\documentclass[a4paper,12pt]{article}
\usepackage{amsfonts,amssymb,amsmath}
\usepackage[mathscr]{eucal}

 \newtheorem{prop}{Proposition}

\begin{document}
\begin{center}

\section*{About the non-standard viewpoint on the dynamics of closed vortex filament}

{ S.V. Talalov}

{Department of Applied Mathematics,
 State University of Tolyatti, \\ 
 14 Belorusskaya str.,  Tolyatti, Samara region,   445020 Russia. \\
 {svt\_19@mail.ru}  }

\begin{abstract}
     In this article   we construct the Hamiltonian description of the closed vortex filament dynamics in terms of non-standard variables, phase space and constraints. The suggested approach   makes    obvious   interpretation of    considered system   as  a quasiparticle  that possess  certain external and internal degrees of the freedom.   The constructed theory is invariant under the transformation of Galilei group.  The appearance of this group allows for a  new viewpoint on the energy of a closed vortex filament with zero thickness. The explicit formula for the effective mass of the quasiparticle
				''closed vortex filament'' is suggested.  
\end{abstract}
\end{center}

{\bf keywords:} closed vortex filaments;  constrained hamiltonian systems;
  effective mass

\section{Introduction}	

  The idea that the closed vortex structures have a particle-like properties is very old\cite{Thom}.
    Although the development of the fundamental particle physics   went in  another direction, the  dynamics of  such structures  continues to attract interest\cite{Moff}.
		When interpreted as  quasiparticles, these structures   may possess some unusual properties.  For instance,  quasiparticle is an object that  demonstrates the anizotropic  reaction on an external force in some cases. Another example is anyons - the ''particles'' with fractional statistics.   As it seems, such objects can be realized as  quasiparticles only. 
		
		The investigation of the vortex dynamics in it's different aspects has a long and rich history. 
		Without attempting  to review  this topic in this short article, let us mention the work \cite{Hasim} where the vortex filament (in the local induction approximation)
		was discribed in terms of solutions of non-linear Schr\"odinger equation.
 It is also relevant to notice  here that 		
		   gauge equivalence     between the non-linear Schr\"odinger equation and the continuous Heisenberg    spin chain exists\cite{TakFad}.

  In this work we  consider the  closed evolving curve ${\boldsymbol{z}}(\tau,\xi)$  that is defined by the formula
\begin{equation}
        \label{involve}
                                   {\boldsymbol{z}}(\tau,\xi) =  {\boldsymbol{z}}_0 + 
         {R_0}\, \int\limits_{0}^{2\pi}  \left[ {\xi - \eta}\right] {\boldsymbol j}(\tau,\eta) d\eta\,.  
                  \end{equation}
                Parameters $\tau$ and $\xi$ are dimensionless parameters.     The constant $R_0$ is  {\it in-put}  constant in our theory which defines the scale of length.
								The notation $[x]$ means the integer part of the number $x/{2\pi}$; the $2\pi$-periodical  vector function ${\boldsymbol j}(\xi)$ satisfies the equation of motion
								
								\begin{equation}
        \label{CHSCeq}
        \partial_\tau {\boldsymbol{j}}(\tau ,\xi) = 
        {\boldsymbol{j}}(\tau ,\xi)\times\partial_\xi^{\,2}{\boldsymbol{j}}(\tau,\xi)\,
        \end{equation}
								             							
                and the consraint equalities:
            \begin{equation}
  \label{constr_j_0}
  {\Phi_k }     =    \int\limits_{0}^{2\pi}{j}_k(\xi)d\xi  = 0\, \qquad  k=1,2,3,
   \end{equation}      
						
						\begin{equation}
  \label{constr_j}
  {\boldsymbol j}^{\,2}(\xi)  = 1\,.
   \end{equation}  
			
			Thus, the field ${\boldsymbol j}(\xi)$  defines  the $2\pi$-periodical continuous Heisenberg    spin chain\cite{TakFad}. Consequently, the  
						curve ${\boldsymbol{z}}(\tau,\xi)$  describes  closed vortex filament 
						that satisfies    the local induction equation:
						
			
			\begin{equation}
        \label{LIE_eq}
        \partial_\tau {\boldsymbol{z}}(\tau ,\xi) = \frac{1}{R_0}\,
        \partial_\xi{\boldsymbol{z}}(\tau ,\xi)\times\partial_\xi^{\,2}{\boldsymbol{z}}(\tau ,\xi)\,.
        \end{equation}
			The   space-time symmetry group in  our theory is the group $E(3)\times  E_\tau$, where  $E(3)$ is the group of motions of space $E_3$ and 
			$E_\tau$ is the group of ''time translations'': $\tau \to \tau +c$.
					        Let us note that the description  of the vortex filament in terms of continuous Heisenberg    spin chain still continues to be interesting
						(see\cite{AbhGuh}, for example).

            \section{Dynamical invariants and  new variables}

  The canonical consideration of the fluid dynamics\cite{Batche} leads to   the following expressions for the momentum and  angular momenta:
   \begin{equation}
        \label{p_and_m_st}
        \boldsymbol{p} = \frac{1}{2}\,\int\,\boldsymbol{r}\times\boldsymbol{\omega}(\boldsymbol{r})dV\,,\qquad 
         \boldsymbol{s} =\frac{1}{3}\, \int\,\boldsymbol{r}\times
         \bigl(\boldsymbol{r}\times\boldsymbol{\omega}(\boldsymbol{r})\bigr)dV\,.
        \end{equation}
       
   The vector  $\boldsymbol{\omega}(\boldsymbol{r})$  means  the vorticity and  the fluid density   $\varrho \equiv 1$  here.                 
      So,   the vorticity for the vortex filament
   \begin{equation}
        \label{vort_w}
     \boldsymbol{\omega}(\boldsymbol{r}) =  \Gamma
                  \int\limits_{0}^{2\pi}\,\hat\delta(\boldsymbol{r} - \boldsymbol{z}(\xi))\partial_\xi{\boldsymbol{z}}(\xi)d\xi\,,
       \end{equation}            
   where the symbol $\Gamma$ denotes the circulation  and the symbol  $\hat\delta(\xi)$ means $2\pi$-periodical $\delta$-function. .  
	Thus, for example, the following formula for the canonical momentum takes place:
	\begin{equation}
        \label{impuls_def}
                 \boldsymbol{p}  =     {R_0}^2 \Gamma                  {\boldsymbol f} \,,
                 \qquad 
                  {\boldsymbol f} = \frac{1}{2}\iint\limits_{0}^{2\pi}  \left[ {\xi - \eta}\right]\,{\boldsymbol j}(\xi)\times{\boldsymbol j}(\eta)d\xi  d\eta\,.
      \end{equation} 
			
	      The similar formula can be written for the angular momenta $\boldsymbol{s}$.
				The corresponding canonical formula for the energy ${\mathcal E}$ gives the unsatisfactory result because of the divergence of the integral.  We will return to this question later.

											Now we intend to construct a hamiltonian dynamical system based on the equation (\ref{LIE_eq}).
          In accordance with our suppositions,  this equation  is the direct consecuence of
        the representation     (\ref{involve}) and the equation (\ref{CHSCeq})   for the function ${\boldsymbol{j}}$.  
							The variable ''velocity of liquid'' is absent in our theory.		
								Moreover we use the dimensionless ''time'' $\tau$ in our model.
								Therefore, the following steps must be done to construct the physically interpreted dynamical system:
								\begin{enumerate}
								      \item  the   formulae for the momentum $\boldsymbol{p}$  and  angular momenta $\boldsymbol{s}$
       must be added   to  the equation (\ref{CHSCeq});
			 \item   the value $\Gamma$   is declared   an additional    dynamical variable in our theory, in relation to   variables $\boldsymbol{j}(\xi)$ and  ${\boldsymbol{z}}_0$.
		We  denote the set of the dynamical variables  
 $\{\,{\boldsymbol{z}}_0\,,\Gamma\,,{\boldsymbol j}(\xi)\,\}$ constrained by the conditions  (\ref{constr_j_0})  and
(\ref{constr_j}) as ${\mathcal A}$. 
        \item the additional dimensional  constant $t_0$ that defines the  scale of time, must be added in our theory.
				\end{enumerate}
				
  The above considerations lead to  the following 
  
					\begin{prop}
										The set ${\mathcal A}$  parametrizes the considered dynamical system - the closed vortex filament
					${\boldsymbol{z}}(\xi) $ 					 evolving in accordance with the equation (\ref{LIE_eq}) and 
					 having a ''position'' (${\boldsymbol{z}}_{01} $,  ${\boldsymbol{z}}_{02}$,  ${\boldsymbol{z}}_{03}$) and circulation
					$\Gamma$.  	This dynamical system has a momentum $\boldsymbol{p}$  and  angular momenta $\boldsymbol{s}$  calculated as prescribed above.
					\end{prop}

    	Next we are going to describe the considered dynamical system in terms of   other  variables.
 The reasons are following:
\begin{itemize}
\item  we suppose that new variables  will allow to interpret our system clearly as a quasiparticle
in a space $E_3$;  
\item we can  expand the symmetry group and use the group-theoretical approach for the  definition of our systems energy;
 \item new variables  will be more suitable for subsequent quantization\footnote{Quantization is beyond the focus of this article.}. 
\end{itemize}
 In fact,   
 we construct the new dynamical system that will be equivalent to the closed  vortex filament in some way.
 To do it let us   reparametrize the set ${\mathcal A}$. 
      First,  let us note that we can use the equivalent variable  $p = |\boldsymbol{p}|$ instead the variable   $ \Gamma $  on the  set ${\mathcal A}$.
			Second, we introduce the spherical coordinates $(p,\theta, \varphi)$ where the $p_3$ axis is parallel to the vector ${\boldsymbol f}$. After that we add the coordinates
			$(\theta, \varphi)$ as additional dynamical variables.
			Third, we replace the spherical variables $(p,\theta, \varphi)$ with the Decart variables $p_1$, $p_2$ and $p_3$ and will use these 
			 quantities  as  new independent fundamental variables. Let us note that corresponding Jakobi determinant $J = p^2\sin\theta$ is degenerated on the $p_3$ axis.
			   This replacement seems physically justified but we introduce three variables instead of one. 
	To avoid the ambiguity in the reconstruction  of the value $ \Gamma $ through the new variables $p_i$ ($i=1,2,3)$,
		 the new set of dynamical variables must be constrained.
		Let us consider the constraint		
			\begin{equation}
  \label{constr_1}
    \Phi_0( p_1, p_2, p_1\,;{\boldsymbol j})\equiv     (\boldsymbol{p} {\boldsymbol f})^2 - \boldsymbol{p}^2 {\boldsymbol f}^2 = 0\,.
     \end{equation}
			
			Of couse, this equality is fullfilled identically if our theory is parametrized by the set ${\mathcal A}$.
We denote the set of the independent variables $({\boldsymbol z_{0}};\, {\boldsymbol p}\,; {\boldsymbol j}(\xi)\,)$  constrained by the conditions (\ref{constr_j_0}),
 (\ref{constr_j}) and  (\ref{constr_1})  as $\Omega$.   			
			The following proposition is true.
\begin{prop}
\label{pr-corresp}
The one-to-one correspondence
$$ {\mathcal A} \quad
 \longleftrightarrow\quad \Omega$$
is valid.
\end{prop}
			The proof follows from the Swarz unequality for the vectors 	$\boldsymbol{p}$ and ${\boldsymbol f}$.	
			 		One constraint	is enough here because the constraint  (\ref{constr_1}) corresponds to the value $\theta =0$  
					where the Jacobi determinant $J=0$.

			The introduction  of the variables  $p_i$    $(i=1,2,3)$ allows to enlarge the   space-time symmetry group $E(3)\times  E_\tau$ of our theory by means of addition
 of Galilei boosts  
       $$ p_j \longrightarrow \tilde{p_j} =  p_j  + cv_j\,,\qquad  c,v_j = const,\quad j=1,2,3\,.$$
			Moreover, we suppose that the procedure of the one-parameter  $(m_0)$  central extension for the standard  Galilei group ${\mathcal G}_3$is fulfilled.
      Consequently,  the symmetry group  for our theory  is an  extended  Galilei group $\widetilde{\mathcal G}_3$. 					
			For convenience, we  introduce  variables
  $$  q_i =  m_0 z_{0i}  + {\tau}t_0 p_i \,,\qquad       i=1,2,3\,, $$ 
  instead of the variables   $z_{0i}$,   $(i=1,2,3)$.  Finally, the variables
                    ${\boldsymbol j}(\xi)$,  ${\boldsymbol{q}}$,  $\boldsymbol{p}$,
                  will be  declared as the new fundamental variables of our model.
 For example, 
     the curve   ${\boldsymbol{z}}(\tau ,\xi)$ must be  reconstructed through these variables in accordance with the  formula
 \begin{equation}
 \label{z_funct}
  {\boldsymbol{z}}(\tau,\xi)   = \frac{1}{m_0}\left({\boldsymbol {q}} - \tau t_0 {\boldsymbol{p}}  \right) + 
         {R_0}\, \int\limits_{0}^{2\pi}  \left[{\xi - \eta}\right] {\boldsymbol j}(\tau,\eta) d\eta\,.
   \end{equation}

			 \section{Energy and hamiltonian structure }

      The extension of the symmetry group makes it possible to suggest an energy definition in our model.  
      			As it is well-known, the straightforward attempt to calculate the energy of an infinite vortex filament by means of canonical formula \cite{Saffm}
       \begin{equation}
  \label{can_energy}
       {\mathcal E} = \frac{1}{8\pi}\,\iint
       \frac{\boldsymbol{\omega}(\boldsymbol{r})\boldsymbol{\omega}(\boldsymbol{r}^{\prime})}{|\,\boldsymbol{r} - \boldsymbol{r}^{\prime}|}dVdV^{\prime}=
       \frac{{\Gamma}^{\,2}}{8\pi}\iint 
       \frac{\partial_\xi{\boldsymbol{z}}(\xi)\partial_\xi{\boldsymbol{z}}(\xi^{\prime})}{|\,{\boldsymbol{z}}(\xi) -  {\boldsymbol{z}}(\xi^{\prime}) |}d\xi d\xi^{\prime}\,,
       \end{equation}
       leads to the unsatisfactory result: the integrals in this formula   diverge.  The standard approach to solve this problem  is to take into account the finite thickness $a$ of the filament
						and  the subsequent  regularization  of  the integral.  The parameter $a$ is absent in our theory. 
						The energy of the arbitrary   configuration  in our model will be   considered     from the group-theoretical viewpoint.
												      Indeed,           the Lee algebra of group $\widetilde{\mathcal G}_3$ has three Cazimir functions:      
           $$ {\hat C}_1 = m_0 {\hat I}\,,\quad 
  {\hat C}_2 = \left({\hat M}_i  - \sum_{k,j=1}^3\epsilon_{ijk}{\hat P}_j {\hat B}_k\right)^2 \,, 
  \quad {\hat C}_3 = \hat H -  ({1}/{2m_0})\sum_{i=1}^3{\hat P}_i^{\,2}\,,$$                         
       where        ${\hat I}$ is the unit operator,     ${\hat M}_i$,   $\hat H$,  ${\hat P}_i$         and  ${\hat B}_i$
        are the respective generators of rotations, time and space translations and Galilean boosts. 
        As it is well known, the function  ${\hat C}_3 $  can be interpreted as  an  ''internal energy of the particle''. 
       			Let us consider the closed filaments that correspond to the points $p_i=0$  ($i=1,2,3$), $j_1(\xi) =\cos\xi$, $j_2(\xi) =\sin\xi$, $j_3(\xi) =0$ (variables $z_i$ take  arbitrary values)
						on the set $\Omega$.
				We postulate here that	the energy of this configuration is equal to the value ${\mathcal E}_0= m_0 R_0^2/t_0^2$.  The constant ${\mathcal E}_0$ defines the energy scale   in our theory.
            Regarding  the arbitrary configurations, 
						we suppose that the Galilei - invariant expression
       $${\hat C}_3 = \frac{{\mathcal E}_0}{2\pi}\,\int\limits_{0}^{2\pi} \bigl( \partial_\xi {\boldsymbol j}(\xi)\bigr)^2 d\xi\,,$$
        is a natural candidate for the internal energy. Of couse,  this expression is motivated by 	
				  the formula for  hamiltonian  of continuous Heisenberg    spin chain.              
       			As a result, the following function on the set $\Omega$ is a good candidate for the energy:                
 \begin{equation}
  \label{energy_1}
 {H}_0(p_1,p_2,p_3\,;{\boldsymbol j}) = \frac{1}{2m_0}\sum_{i=1}^3 p_i^{\,2}   +   \frac{{\mathcal E}_0}{2\pi}\,\int\limits_{0}^{2\pi} \bigl( \partial_\xi {\boldsymbol j}(\xi)\bigr)^2 d\xi\,.
\end{equation}                
  To complete the consideration of energy, we must define the Poisson brackets that are  compatible with the dynamics and constraints.  
	 Pursuant to Dirac prescriptions about the primacy of  Hamiltonian structure,
  we define such structure  axiomatically here.  The correspondent definitions are following.
 
  \begin{itemize}
  \item Phase space ${\mathcal H} = {\mathcal H}_j \times {\mathcal H}_3 $. The space $ {\mathcal H}_3$ which is parametrized by the variables 
   ${\boldsymbol{q}}$ and  ${\boldsymbol{p}}$ is the phase space of a free structureless $3D$ particle. The space    $ {\mathcal H}_j$ is parametrized by the $2\pi$-periodical
	functions    ${j}_k(\xi)$, where $k=1,2,3$.

   \item Poisson structure:
  \begin{eqnarray}
  \{p_i\,,q_j\} & = & m_0\,\delta_{ij}\,,\qquad i,j =1,2,3\,, \nonumber \\
  \label{ja_jb}
  \{ j_a(\xi), j_b(\eta)\} & = & \beta\, \epsilon_{abc} j_c(\xi) \hat\delta(\xi- \eta)\,, \qquad      \epsilon_{123} = 1\,.
  \end{eqnarray}
  where $\beta =  - 2/{\mathcal E}_0 t_0 $.  
	All other brackets  vanish. 
	In accordance with the definition (\ref{ja_jb}), the function  ${\boldsymbol{j}}^{\,2}(\eta)$ annulates the brackets of the fundamental variables. Thus the condition
   (\ref{constr_j}) selects the symplectic sheet in the phase space  $ {\mathcal H}_j$;
  \item  Constraints:   $\Phi_k=0$, where $k=0,\dots,3$. The functions $\Phi_k$ were defined in the eq. (\ref{constr_j_0})  and  (\ref{constr_1});
  \item Hamiltonian 
	$$H=H_0+ \sum_{k=0}^3 l_k\Phi_k\,,$$
	where  the function  $H_0  $ was defined by the formula (\ref{energy_1})
    and the values ${l}_k$ are the Lagrange factors.
  \end{itemize}

\begin{prop}
The constraints system   (\ref{constr_j_0})  and  (\ref{constr_1}) is defined  first type constraints in the Dirac terminology.
\end{prop}
Indeed, the formulae
$$  \{ \Phi_a, \Phi_b  \}  = \beta\, \epsilon_{abc} \Phi_c\,, \qquad   a,b,c = 1,2,3 $$
are the direct consequence of the definition (\ref{ja_jb}).
 Let us calculate the brackets  $\{ \Phi_a, \Phi_0  \} $.
So, the straightforward  calculations lead to the following brackets:
$$  \{ \Phi_a, f_k  \}  = -\beta \int\limits_{0}^{2\pi}  \left[{\xi - \eta}\right] {j}_k(\xi){j}_a(\eta) d\xi d\eta\,,
\qquad a,k = 1,2,3\,,$$
that are true if the constraints (\ref{constr_j_0}) were are fulfilled.
Consequently, 
\begin{eqnarray}
~&~&~\{ \Phi_a, \Phi_0  \}   =  2\beta p^2 \int\limits_{0}^{2\pi}  \left[{\xi - \eta}\right] 
\bigl({\boldsymbol f}{\boldsymbol j}(\xi)\bigr){j}_a(\eta) d\xi d\eta  - \nonumber\\
~~~& - & \beta\bigl({\boldsymbol p}{\boldsymbol f}\bigr) \int\limits_{0}^{2\pi}  \left[{\xi - \eta}\right]
 \bigl({\boldsymbol p}{\boldsymbol j}(\xi)\bigr){j}_a(\eta) d\xi d\eta  -
\beta\bigl({\boldsymbol p}{\boldsymbol f}\bigr) \int\limits_{0}^{2\pi}  \left[{\xi - \eta}\right] 
\bigl({\boldsymbol p}{\boldsymbol j}(\xi)\bigr){j}_a(\eta) d\xi d\eta \nonumber\,.
\end{eqnarray}
Because  the equality ${\boldsymbol p} = const\cdot{\boldsymbol f}$  is fulfilled on the constraint surface  (\ref{constr_1}), the r.h.s. of this formula is vanished   if the equality $\Phi_0 = 0$ holds.

We can easyly prove that 
the constraint surface $\Omega$  agrees  with the dynamics:
the equalities
$\{ H, \Phi_k\} =0$   are fulfilled  on the symplectic sheet  (\ref{constr_j})  if the conditions
(\ref{constr_j_0})  and  (\ref{constr_1})  are fulfilled. 
This fact means that there are no additional constraints in our theory .

Let us define the ''time'' $t=t_0\tau$. The  following equation is true:
$$\frac{d {\boldsymbol{z}}(\xi) }{d t}= \frac{\partial  {\boldsymbol{z}}(\xi) }{\partial t} + \{H_0, {\boldsymbol{z}}(\xi)\}\,.$$
The  calculations of the derivatives here imply that the arguments of the function ${\boldsymbol{z}}$ are the time $t$ and the fundamental coordinates
 ${\boldsymbol{q}}\,,   {\boldsymbol{p}}\,, {j}_a(\xi) $ here.
The detail calculations with the help of the explicit representation   (\ref{z_funct}) give the final expression for the r.h.s. of the last equality in the form
$$(1/{t_0R_0})\,
        \partial_\xi{\boldsymbol{z}}(\xi)\times\partial_\xi^{\,2}{\boldsymbol{z}}(\xi)\,.$$

As a result, we have the following
\begin{prop}
The introduced hamiltonian structure defines  the hamiltonian flows on the set $\Omega$  correctly. The flow that corresponds to  the values ${l}_k =0$, $k=0,\dots,3$, leads to the dynamical equation 
 (\ref{LIE_eq}) for the curve   ${\boldsymbol{z}}(\tau ,\xi)$ that reconstructed through the coordinates of the space ${\mathcal H}$ in accordance with formula (\ref{z_funct}).
   \end{prop}

As usual, we may  substitute  all constraints  explicitly after  calculating all the Poisson brackets.  Therefore, 
the natural expression  for  the energy of our system is:

$${\mathcal E} = H\Big\vert_\Omega   =  \frac{1}{2m_0}\,\Bigl(\boldsymbol{p}\,\boldsymbol{n}_{\!f}\Bigr)^2 + 
  \frac{{\mathcal E}_0}{2\pi}\,\int\limits_{0}^{2\pi} \bigl( \partial_\xi {\boldsymbol j}(\xi)\bigr)^2 d\xi\,,$$
 
where vector $\boldsymbol{n}_{\!f}  = \boldsymbol{f}/|\boldsymbol{f}|$.
This formula leads to the following expression for the inverse effective mass   tensor $(1/ m_{\rm eff})_{ik}$ :
$$\left(\frac{1}{m_{\rm eff}}\right)_{ik}  \equiv  \frac{\partial\mathcal E}{{\partial p_i}{\partial p_k}} = 
 \frac{1}{m_0}(\boldsymbol{n}_{\!f})_i (\boldsymbol{n}_{\!f})_k\,. $$

The model  has three dimensional constants:  $R_0$, $m_0$ and  $t_0 $. 
                These constants define the scale of  length,  mass and   time in our theory.

 \section{Concluding remarks }
Among other directions,
the important  investigations of the closed vortex filament are connected  to studying  its response to external forces.
The experimental studies (see\,\cite{Kop,Pim}, for example) lead to some arguments that this response  may be anisotropic. 
The author hopes that the presented studies have provided further theoretical arguments for such behavior of the vortices.


\begin{thebibliography}{0}
\bibitem{Thom}{W. Thomson, {\it Phil. Mag.} {\bf 34}   (1867) 15–24.}

\bibitem{Moff}{K. Moffatt, {\it Rus. J. Nonlin. Dynamics} {\bf 2}  (2006) 401–410. }

\bibitem{Hasim}{H. Hasimoto,  {\it  J. Fluid Mech.} {\bf 23} (1972)  61 -- 97. }

\bibitem{TakFad}{L.A. Takhtajan, L.D. Faddeev,  {\it Hamiltonian Methods in the Theory of Solitons.}   (Springer-Verlag, Berlin, 1987)}

\bibitem{AbhGuh}{K. Abhinava,  P. Guhaby, {\it Inhomogeneous Heisenberg Spin Chain and Quantum
Vortex Filament as Non-Holonomically Deformed NLS} arXiv: math-ph/1703.02353v2 (2017).}

 \bibitem{Batche}{G.K. Batchelor, {\it An Introducton to Fluid Dynamics} (Cambrige Univ. Press, 1970).}

  \bibitem{Saffm}{P.G. Saffman, {\it Vortex dynamics} (Cambrige Univ. Press, 1992). }
	
	\bibitem{Kop}{V. F. Kopiev, N. N. Ostrikov. Corona discharge microjets  as a possible actuators for jet noise control. {\it TsAGI Science Journal.}
  {\bf 41}.   No1.  97-107  (2010).}
	
	\bibitem{Pim}{V.G. Pimshtein.  Direct observation of reaction and scattering of sound on vortices and radiation of sound by vortices in subtonoc  turbulent jets at aeroacoustic interaction.{\it TsAGI Science Journal.}    {\bf 43}.   No 6.  813 -- 820. (2012). }
         \end{thebibliography}
    \end{document}